\documentclass{aa}
\usepackage{txfonts}
\usepackage{graphicx}

\begin{document}


\title{A Submillimetre Selected Quasar in the Field of \object{Abell478}
       \thanks{Based on observations made with ESO Telescopes at the 
Paranal Observatory under programme IDs 63.O-0087 and 68.A-0111. Also
       based on observations with ISO, an ESA project with instruments
       funded by ESA member states (especially the PI countries: the
       Netherlands, the United Kingdom, Germany, and France) with the
       participation of ISAS and NASA.}}

\author{K.K.~Knudsen
   \and P.P.~van~der~Werf
   \and W.~Jaffe}

\institute{Leiden Observatory, P.O.\ Box 9513, NL--2300 RA Leiden, The Netherlands \\ email: [kraiberg,pvdwerf,jaffe]@strw.leidenuniv.nl}

\offprints{K.\ K.\ Knudsen, \email{kraiberg@strw.leidenuniv.nl}}

\date{Received / Accepted }

\abstract{ 
We present the discovery of a dusty quasar, SMM\,J04135+10277, 
located behind the 
galaxy cluster \object{Abell 478}. The quasar was discovered as the
brightest source in a submillimetre
survey of high redshift galaxies lensed by foreground
rich clusters of galaxies in a project aimed at studying the cosmic star 
formation history of dusty galaxies. With
submillimetre fluxes of 
$S_{850} = 25\pm 2.8\ {\rm mJy}$ and $S_{450} = 55 \pm 17\ {\rm mJy}$
this object is one of the brightest submillimetre sources known. 
Optical imaging revealed a point source with
$I= 19.4\pm0.1\ {\rm mag}$ (corrected for galactic extinction). Follow-up 
optical spectroscopy showed this object to be a quasar at 
redshift $z=2.837\pm 0.003$.
The quasar was also detected at shorter infrared wavelengths
with the Infrared Space Observatory. 
This object is the first quasar discovered by its submillimetre
emission. Given the general lack of overlap between deep submillimetre
and X-ray samples, usually interpreted as a low incidence of active nuclei in
submillimetre samples, this is an unusual object. Analysis of number
counts of quasars and of submillimetre galaxies bears out this suggestion.
We compare the properties of SMM\,J04135+10277 to those
of optically selected quasars with submillimetre emission, and argue
that the optical faintness results from a large viewing angle with the
direction of relativistic beaming, and not from abnormally high
extinction. We also find indications that the bulk of the submillimetre
flux density is not powered by the quasar nucleus. This conclusion is
supported by analysis of the infrared spectral energy distribution.
These results are consistent with previous 
observations that quasars at 
higher redshift tend to have a more prominent cold dust component, most
likely powered by extended star formation in the host galaxy.
{The temperature for the cold dust component is found be $T = 29\pm2\,{\rm K}$
when assuming $\beta=1.5$ for a modified blackbody.}
The quasar is found to have a total infrared luminosity of $(2.9\pm 0.5)\cdot
10^{13}\,{\rm L_{\odot}}$, dominated by the emission from cool dust.

}

\maketitle


\section{Introduction}

Major advances in submillimetre (submm) continuum observations came with 
the Submillimetre Common-User Bolometer Array 
(SCUBA, Holland et al.\ \cite{Holland}), which 
is mounted at the 15\,m James Clerk Maxwell Telescope (JCMT) 
at Mauna Kea, Hawaii.  
This dual-channel instrument for the first time allowed sensitive
mapping, making it possible 
to survey larger areas of the sky to greater depths
than previously possible at submm 
wavelengths.  This development led to the discovery of a new class of objects
of high infrared (IR) 
luminosity, located at cosmological distances (e.g., Smail et
al. \cite{Smail}).
Even though these objects are less common than Lyman-break galaxies 
at similar redshifts, 
they would dominate the cosmic star formation rate density at these
redshifts, if star formation is indeed the source of their high luminosity
(e.g., Blain et al. \cite{Blainetal99}). Since the importance of these
objects was realized,
a number of submm surveys have been performed or are in 
progress (e.g. Smail et al. \cite{SIBK} ; Eales et al.\ \cite{Eales};
Scott et al.\ \cite{scott}; Chapman et al.\ \cite{chapman_lens};
Knudsen et al.\ in prep.).  
One of the biggest
challenges for those surveys has turned out to be the follow-up
observations and the identification of the counterparts causing the 
submm emission. As a result, the determination
of the nature and redshift of these objects has been significantly
hampered.
While currently more than a hundred submm sources have been detected, 
less than 20 of these have reliably been identified with sources 
at other wavelengths, in spite of painstaking attempts.  
The number of published spectroscopic redshifts is even significantly
smaller. The 
majority of the reliably identified counterparts are very or extremely 
red objects (e.g., Frayer et al.\ \cite{smmj00266}; Smail et al.\ 
\cite{smail_smmero}); 
several are also exhibiting active galactic nuclei (AGN) 
features in their spectra (Ivison et al.\ \cite{smmj02399}). 
Because of the small number of secure identifications, any new identification 
adds important information to our understanding of the submm population. 
A survey with different selection criteria, radio-preselected and 
submm-detected, has produced 10 spectroscopic 
redshifts in a sample of 34 sources (Chapman et al.\ \cite{chapman_nat}). 

We are carrying out an extensive SCUBA survey
of a number of galaxy clusters fields, aimed at detecting
gravitationally amplified background galaxies: the Leiden-SCUBA Lensed
Survey (Knudsen et al.\ {\it in prep.}).
In the course of doing the optical
identifications and follow-up of this survey we discovered one of our
submm sources to be a previously unknown type-1 quasar 
(previously reported in Knudsen et al.\ \cite{knudsen00}).  
While submm surveys of optically selected quasars have been
quite succesful (Isaak et al. \cite{Isaak}), this object is the
first type-1 quasar first discovered by its submm emission.
In contrast, type-2 quasars have been detected in small 
numbers in other submillimetre surveys (e.g.\ \object{SMM\,J02399-0136},
Vernet \& Cimatti \cite{vernet}), and
IRAS-radio-optical quasars have been selected before at a wide
range of redshifts (e.g.\ \object{APM\,08279+5255} in Irwin et al.\ 
\cite{irwin}).
In this paper we present the observations of the quasar.
We discuss unusual properties of the object, its optical spectrum, and its IR 
spectral energy distribution, and compare the
results to optically selected quasars. 
We adopt an $\Omega_0 = 0.3$ and $\Lambda = 0.7$ cosmology with 
$H_0 = 70\ {\rm km\,s}^{-1}{\rm Mpc}^{-1}$. 


\section{Observations and results}
\label{sect:obs}

\subsection{Submillimetre data}

The SCUBA data of the $z=0.088$ galaxy
cluster \object{Abell 478} have been obtained during five nights 
in September and December 1997, March 1998 and December 1999. 
The first data were obtained in a program
to study the cooling flow in the cluster itself. In these
data a bright point source was detected. Consequently, extra 
data was obtained to study this object better. 
The total integration time was 6.6 hours (excluding overheads), recording 
data at both 850\,$\mu {\rm m}$ and 450\,$\mu {\rm m}$ simultaneously in
jiggle-map mode. 
The data were obtained mostly under good conditions with  
850\,$\mu {\rm m}$ zenith atmospheric opacity typically around 0.2. 
The pointing was checked regularly and was found to be stable.  
Calibration maps of CRL618 were also obtained. 
The data were reduced using the {\sc SURF} (SCUBA User Reduction 
Facility) and {\sc KAPPA} software packages 
(Jenness \& Lightfoot \cite{surf}). The resulting images have an angular
resolution of $15''$ at $850\,\mu$m and $8''$ at $450\,\mu$m.

Source extraction and estimation of the uncertainties
were carried out using a method based on Mexican Hat 
wavelets (Cay\'on et al.\ \cite{mexhat}; Barnard et al.\ in prep; 
Knudsen et al.\ in prep), which was adopted for the
entire Leiden-SCUBA Lensed Survey, and which will be 
described in a forthcoming publication (Knudsen et al.\ in prep.), 
where the full survey will be presented. This method was
adopted because it is mathematically rigorous and its performance on
SCUBA jiggle maps can be fully
characterized.
Monte Carlo simulations have been performed to determine 
the noise and uncertainties of the derived parameters.  
The area-weighted noiselevels of the maps are 2 mJy at 
$850\ \mu{\rm m}$ and 14 mJy at $450\ \mu{\rm m}$. 

In the $850\,\mu{\rm m}$ map four sources were detected of which the 
brightest has a flux of $S_{850} = 25\pm 2.8\ {\rm mJy}$.  This is 
the only source in the map with detected $450\,\mu{\rm m}$ emission, 
$S_{450} = 55\pm 17\ {\rm mJy}$.  It was detected with a signal-to-noise
of 15, for which the formal positional uncertainty including the pointing 
uncertainty of the JCMT is $3.2''$.   This is the object 
\object{SMM\,J04135+10277} for which we 
are here presenting the follow-up observations.  Fluxes and positions
are presented in Table~\ref{tab:flux_pos}.

\subsection{Optical identification}

\begin{figure}
\resizebox{\hsize}{!}{\includegraphics{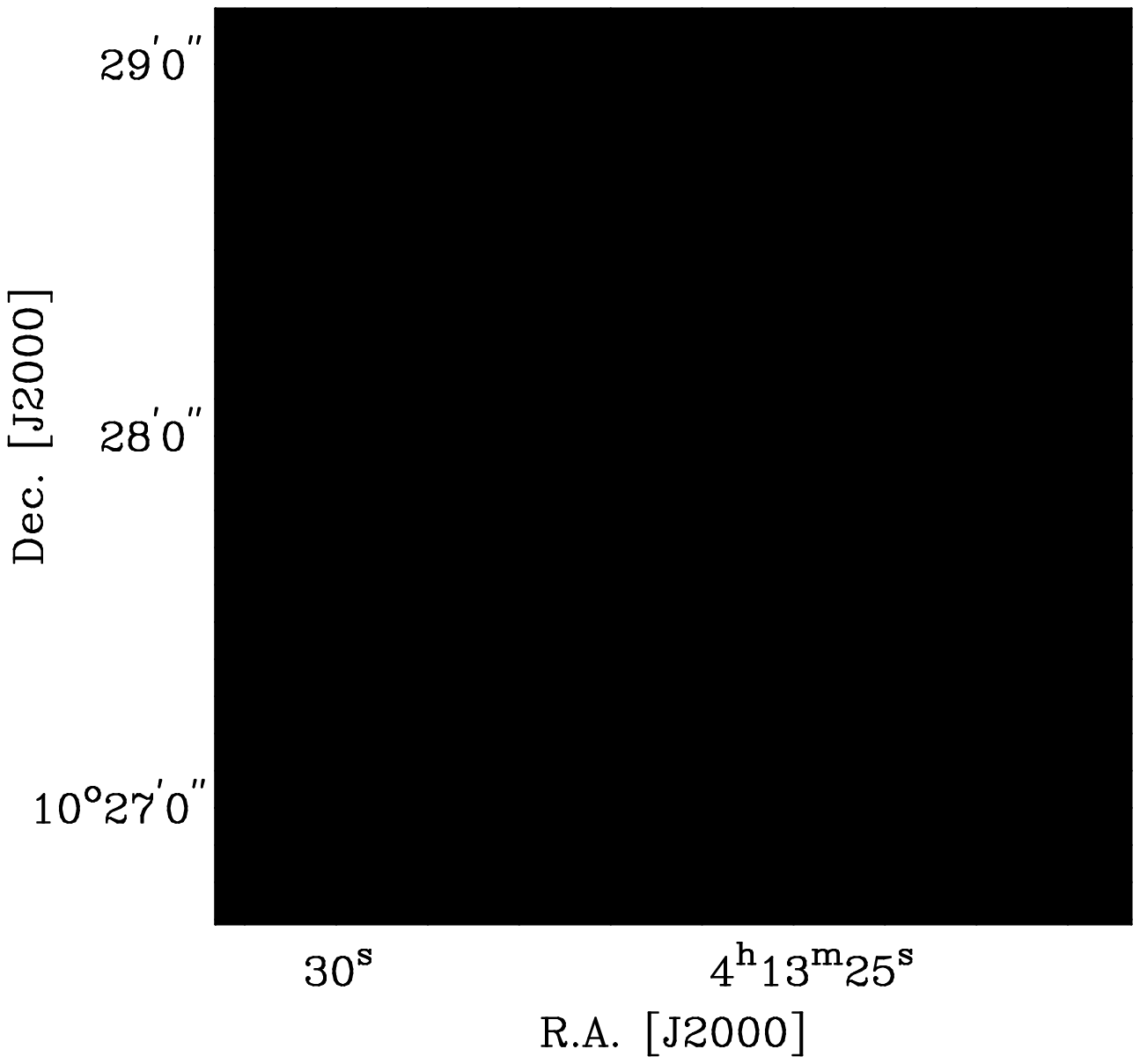}}
\center{\resizebox{6cm}{!}{\includegraphics{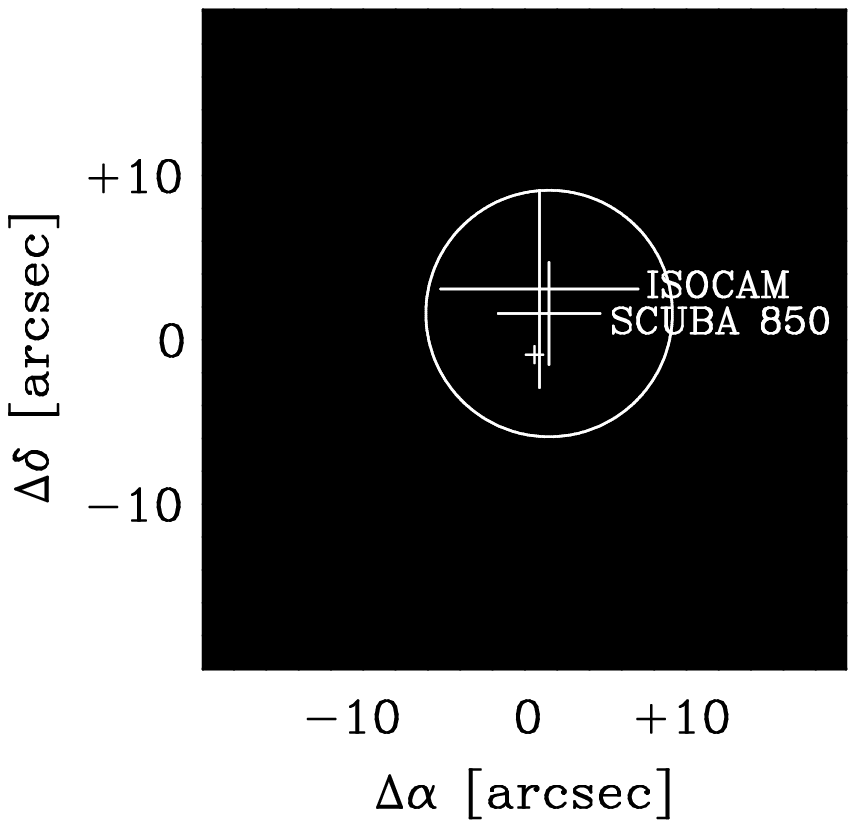}}}
\caption{{\it Top:}\ VLT FORS1 $I$-band image of A478 overlayed with 
the contours of the SCUBA $850\,\mu{\rm m}$ map.  
The contours represent the $850\,\mu{\rm m}$
signal-to-noise ratios of 3,4,5,6,8,10,12,14 -- $1\sigma = 2\,{\rm mJy}$.
{\it Bottom:}\ Zoom in on the quasar.  This box is centered on the 
optical position of the quasar and has a size of $40''\times 40''$. 
The circle shows the size of the
SCUBA $850\,\mu{\rm m}$ beam centered on the SCUBA detection.  The biggest cross
indicates the position in the ISOCAM $14.3\,\mu{\rm m}$ map, where the total 
astrometric uncertainty is $6''$.  The medium cross indicates the detection 
position at $850\,\mu{\rm m}$, the positional uncertainty is $3.2''$.  The 
smallest cross is the radio position. 
}
\label{fig:FORS1SCUBA}
\end{figure}

For identification and redshift determination of the SCUBA source(s) 
optical imaging and spectroscopy was obtained with FORS1 at VLT-UT1
(Antu) in Chile, 
in September 1999. 
Four 15 min exposures in $I-$band were aquired in photometric conditions. 
The frames were bias-subtracted, flatfielded and stacked.  The resulting
image is 
shown in Fig.~\ref{fig:FORS1SCUBA} with the SCUBA $850\ \mu{\rm m}$ 
contours overlayed. 
The seeing measured in the final image is $0\farcs 9$. 
The standard star field \object{PG0231+051} (Landolt \cite{landolt}) 
was used for the calibration.  
The source detection and photometry was performed using 
SExtractor (Bertin \& Arnouts \cite{SEx}). The center of 
\object{SMM\,J04135+10277} is coincident
with an $I=20.5\pm 0.1\ {\rm mag}$ point source at 
$\alpha = 04^{\rm h}13^{\rm m}27\fs28$, $\delta = 10\degr 27\arcmin40\farcs4$ 
(J2000).  There are no other apparent candidate counterparts. 
The optical position is within the error circle of the submm observation.
One of the other SCUBA sources (\object{SMM\,J04134+10270}) 
coincides with a galaxy, which, given its size and magnitude, is a probable 
cluster member.  There are no obvious candidate counterparts for the two 
other SCUBA sources. 
Using the DIRBE/FIRAS maps (Schlegel et al.\ \cite{dustmaps}), a
Galactic reddening $E(B-V) \approx 0.52\ {\rm mag}$ is derived --- a 
substantial reddening.  The corrected
$I$ magnitude of the optical counterpart for SMM\,J04135+10277 is thus 
$19.4\pm 0.1\ {\rm mag}$.  

\subsection{Optical spectroscopy}

FORS1 spectroscopy of SCUBA sources in the A478 field was also obtained
in September 1999. We used FORS1 in Multi-Object Spectroscopy (MOS) mode
to obtain spectra of a number of targets, using grism 150I+17, without
order sorting filter. This setup gives a spectral resolution of 260 at
720\,nm with the $1''$ slit which we employed. 
Overlap of the second spectral order may affect the wavelength
region longwards of 650\,nm, but was in the present case found not to
affect the spectra.
Two exposures of 1800\,sec were obtained in a seeing of $1.3''$.
The spectra were bias-subtracted and flatfielded. Wavelength calibration
was achieved using exposures of He and Ar lamps. Correction for telluric
absorption and flux calibration was carried out using observations of
the white dwarf EG274 ($V=11.03$), which we corrected for photospheric
absorption features. The multislit mask included both the bright SCUBA source
SMM\,J04135+10277 and the fainter SCUBA source SMM\,J04134+10270. 
The extracted spectra were corrected for galactic foreground absorption
using the DIRBE/FIRAS maps.

The optical spectrum of \object{SMM\,J04135+10277} (Fig.~\ref{fig:spectrum})
shows broad emission lines, 
of which the four most prominent can be identified with Ly$\alpha$+N\,{\sc v}, 
Si\,{\sc iv}+O\,{\sc iv}], C\,{\sc iv} and C\,{\sc iii}. In addition the
spectrum shows a power-law continuum. All of these features are
characteristic of quasars. 
Bluewards of the Ly\,$\alpha$ emission line, Ly\,$\alpha$ 
forest absorption is seen. 
We use the C\,{\sc iii}, C\,{\sc iv} and the Si\,{\sc iv}+O\,{\sc iv}] 
lines, with the largest 
weight on the symmetric C\,{\sc iii} line, to determine the redshift. 
We find the value $z = 2.837 \pm 0.003$, consistently for the peak value
of all three profiles.  

The spectrum of \object{SMM\,J04134+10270} 
confirms its membership of the A478 cluster. 
It shows the characteristic spectrum of a quiescent elliptical galaxy
with no evidence for nuclear activity.  This source will be discussed 
together with the rest of the survey in Knudsen et al.\ (in prep.).

\subsection{Near-infrared spectroscopy}

Since restframe ultraviolet emission lines of quasars can be
significantly blueshifted with respect to the systemic velocity
(e.g., Carswell et al.\ \cite{Carswell}), 
we also attempted to obtain
additional redshift information using restframe optical
lines. Unfortunately, at $z\approx2.84$, the brightest lines (H$\alpha$,
H$\beta$, [O\,{\sc iii}] 5007\,\AA) are all in wavelength regions where the
earth atmosphere is opaque. We therefore concentrated on the [O\,{\sc ii}]
3727\,\AA\ line which is redshifted to the blue edge of the H-band
window, a region strongly affected by atmospheric absorption lines. 

We used ISAAC on VLT-UT1 (Antu) in February 2002, to take H-band
spectra of \object{SMM\,J04135+10277}. We used the medium resolution grating
with a $1''$ slit to obtain an $R=3000$ spectrum between 1.41 and
$1.49\,\mu$m, which should contain the [O\,{\sc ii}] line for redshifts between
2.78 and 3.00. In addition, we obtained a low-resolution spectrum
($R=500$) of the entire H-band, in an attempt to detect H$\gamma$, which
although intrinsically faint, should at least lie in a clear part of the
spectrum. Both spectra were obtained in photometric conditions and in an
optical seeing of $0.8''$, by nodding the object along the
slit. Total integration time was 44\,minutes in both spectra. The
individual frames were pairwise subtracted in order to remove the bright
OH nightsky lines, flatfielded and coadded. Wavelength calibration was
derived from the OH nightsky lines. Correction for telluric absorption
and flux calibration were achieved using 
the B5V star Hip25499 (H=5.62) and the B2V star Hip28142 
(H=7.497), corrected for
photospheric absorption. Unfortunately, 
while the continuum of the quasar is clearly
detected in both spectra, no emission features are seen. Undoubtedly,
this is due to atmospheric absorptions in the region of the redshifted
[O\,{\sc ii}] line, and the faintness of the relevant features in the rest
of the H-band spectrum.

\begin{table}
\caption[]{Coordinates and fluxes at different wavelengths for 
SMMJ04135+10277.  The fluxes as they are listed here have not been 
corrected for the gravitational lensing.
\label{tab:flux_pos}}
\begin{tabular}{llll}
\hline
\hline
passband & RA(J2000) & Dec(J2000) & $f_{\nu}$ \\
\hline
$850\mu{\rm m}$    & 04:13:27.2  & +10:27:43   & $25\pm2.8\, {\rm mJy}$ \\
$450\mu{\rm m}$    & 04:13:27.2  & +10:27:42   & $55\pm17\, {\rm mJy}$ \\
$180\mu{\rm m}$    &             &             & $< 620\, {\rm mJy}$  \\
$14.3\mu{\rm m}$   & 04:13:27.24 & +10:27:44.5 & $470\pm80\, {\rm mJy}$ \\
$6.7\mu{\rm m}$    & 04:13:27.88 & +10:27:43   & $200\pm30\, {\rm mJy}$ \\
$I$                & 04:13:27.28 & +10:27:41.4 & $19.4\pm 0.1\, {\rm mag}$ \\
$4.86\, {\rm GHz}$ & 04:13:27.26 & +10:27:40.5 & $220\pm35\, \mu{\rm Jy}$ \\
$1.4\, {\rm GHz}$  &             &             & $< 750\, \mu{\rm Jy}$ \\
\hline
\end{tabular}
\end{table}

\subsection{ISO data}

We also inspected the archive of the Infrared Space Observatory (ISO,
Kessler et al.\ \cite{ISO})
and extracted observations of A478 using both the mid-infrared 
camera (CAM, Cesarsky et al.\ \cite{CAM}) 
and the mid/far-infrared photometer (PHOT, Lemke et al.\
\cite{PHOT}).

The PHOT data were obtained using the P22 raster mode at $180\,\mu$m,
with $92''$ pixels, on February 21, 1998. 
The data were
reduced using the PHOT Interactive Analysis (Gabriel et al.\ \cite{PIA}). 
Initial data reduction steps
included discarding of corrupted data, non-linearity correction, and
deglitching of individual ramps. After fitting all integration ramps
with a first order polynomial, further deglitching and data editing, 
and dark current subtraction, the data were corrected for detector
drifts and for vignetting, and calibrated using the internal Fine
Calibration Sources. The resulting image shows a $0.62\pm0.19\,$Jy
source, the centroid of which is however
displaced by $46''$ from SMM\,J04135+10277. This displacement is
less than the ISO angular resolution at $180\,\mu$m, but
much more than
the nominal ISO pointing uncertainty of $2''$. While the centroid of a faint
source can be displaced somewhat when placed on top of a highly
structured background, in the present case the offset is so large that
the $180\,\mu$m detection cannot reliably be associated with the
quasar. Hence in the following we label this as
an upper limit.

The CAM data were obtained using the LW3 filter (effective wavelength
$14.3\,\mu$m) on February 21, 1998, 
and using the LW2 filter (effective wavelength
$6.7\,\mu$m) on March 21, 1998 using the CAM01 raster observing mode
with $6''$ pixels. The data were reduced using the CAM Interactive
Analysis package (Ott et al.\ \cite{CIA}). 
Processing steps consisted of dark current
subtraction, deglitching and correction of transients using the PRETI
method (Aussel et al.\ \cite{Aussel}), 
which is particularly suited for the detection of faint
sources, flatfielding using a flatfield derived from the stacked
dataframes, and mosaic contruction taking into account the image
distortion. This resulted in clear detections of SMM\,J04135+10277 at
$14.3\,\mu$m with a flux density of $0.47\pm0.08\,$Jy, and at
$6.7\,\mu$m with a flux density of $0.20\pm0.03\,$Jy. Positions of
these sources are listed in Table~\ref{tab:flux_pos}. 


\subsection{CO $J=3{\to}2$ emission}

In a recent commisioning project of the new COBRA spectrometer on
  the Owens Vally Radio Observatory, the 
CO $J=3{\to}2$ emission line has been detected from the quasar. This
detection confirms the nature of 
\object{SMM\,J04135+10277} as a hyperluminous IR 
quasar. The redshift implied by the CO line is $z=2.84$, which is indeed
  somewhat higher than the optically determined redshift.
 This result will be discussed in detail in 
Hainline et al.\ (in prep).


\section{Discussion}
\label{sect:res_disc}

\subsection{SMM\,J04135+10277 and the optical quasar population}

Only little is known about the importance of AGNs in the submm 
population.  Most studies comparing X-ray and submm observations 
conclude that the submm population is powered by star formation rather 
than AGNs and especially quasars (e.g., Almaini et al.\ \cite{almaini}),
based on the lack of overlap of X-ray and submm sources in deep studies.
This has been confirmed in a study combining very deep 
{\it Chandra} observations with SCUBA observations of the HDF-N, 
where Alexander et al.\ (\cite{alexander}) found that a significant 
fraction of bright submm sources ($f_{850\mu{\rm m}}>5\ {\rm mJy}$)
harbour an AGN, however, the AGN is not powerful enough to power the 
submm emission.  
This makes SMM\,J04135+10277 a particularly interesting object, since
here we have a bright submm source that is unequivocally identified with
a type-1 quasar. 
Given what is known about the abundance of type-1 quasars, is 
this an ordinary object that we should have expected to 
find in our survey, or are we dealing with an exceptional case?  We here 
estimate the probability of finding a high
redshift submm emitting quasar in our survey.
The total area of our survey is $65\ {\rm arcmin}^2$ (Knudsen et al., 
in prep.).  
Using the optical spectrum we estimate that the quasar has a $B$ magnitude
$B \sim 21.0-21.5\ {\rm mag}$.  Based on 
the counts of Kennefick et al.\ (\cite{Kennefick}), we find that there
is only a 20\% probability of finding a quasar with 
$z > 2.3$ and $16.5\,{\rm mag} < B < 22\,{\rm mag}$ in our
survey. Furthermore, the probability that such a quasar is a bright submm
source is also less than unity, as shown by 
Priddey et al.\ (\cite{Priddey}), 
who did a submm study of optically selected 
quasars at $1.5 < z < 3$.  For the sub-sample of quasars with 
$z>2.3$, only 30\% of these had detectable submm emission down to 6.8 mJy
and all of those are fainter than \object{SMM\,J04135+10277}.  
Combining the numbers
we estimate only a 6\% chance of detecting a submm bright quasar at
$z>2.3$ in our survey, if that quasar was drawn from the population of
optically selected quasars. 

We also estimate the expected number of bright submm sources in 
the surveyed area, regardless of their physical nature.  
According to the number counts from Smail et al.\ 
(\cite{SIBK}) we should expect to find two sources with $850\,\mu$m
fluxes between 
20 and 25\,mJy.  Our observations (Knudsen et al., in prep.) are 
in agreement with that number.  Comparing this to the small chance of finding 
a high redshift submm emitting quasar in our survey, this result suggests that 
the bright part of the submm population does not originate from dusty
quasars, and that \object{SMM\,J04135+10277} is an unusual object.

\subsection{Optical spectrum}
\label{subsect:opt_spec}

Since \object{SMM\,J04135+10277} 
is the first quasar selected based on its submm
emission, it is of interest to compare its properties to those of
optically selected quasars.
Turning first to the optical spectrum,
the shapes of the C\,{\sc iii}, C\,{\sc iv} and the Si\,{\sc iv}+O\,{\sc iv}] 
lines appear as expected.  However,
the Ly\,$\alpha$+N\,{\sc v} emission line has a more unexpected shape.  The 
peak and blue wing appear to be absorbed.  Furthermore, the strength of the 
line relative to the other emission lines is unusually low for a
quasar. Since dust is
present in this quasar, it is natural to assume that atomic hydrogen
will also be present, so that associated
absorption may play a role in suppressing the Ly\,$\alpha$ emission.
However, for a 
more detailed assessment of this effect, 
a higher resolution spectrum is needed. 
Comparing the optical spectrum of SMM\,J04135+10277 to that of quasars 
selected at other wavelengths (see
e.g. Francis et al.\ (\cite{Fran92}) for a composite spectrum), there are no 
significant differences except for the suppressed Ly\,$\alpha$ emission.

\begin{figure}
\resizebox{\hsize}{!}{\includegraphics{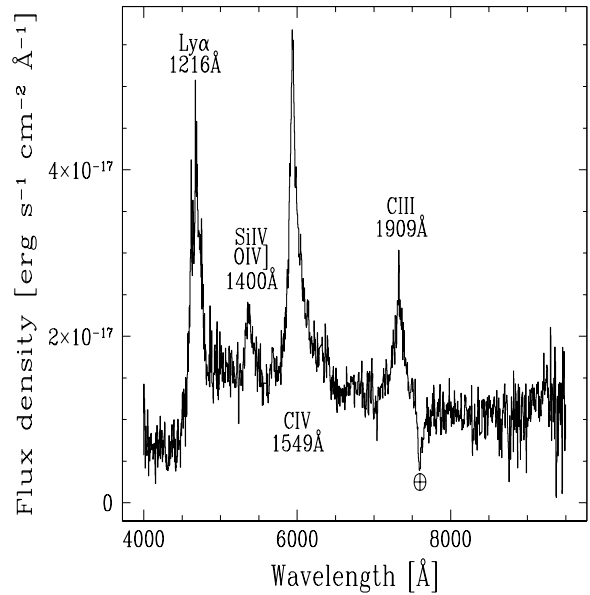}}
\caption{FORS1 spectrum of the SMM\,J04135+10277.  
The Earth-symbol indicates a telluric absorption line.  
The spectrum has been corrected for Galactic extinction.}
\label{fig:spectrum}
\end{figure}

We note that SMM\,J04135+10277 belongs to the optically fainter part 
of the quasar population. Can this be the effect of strong 
absorption by dust, which would
then simultaneously account for the luminous dust emission from
SMM\,J04135+10277? 
We obtain a measure of the isotropic
luminosity of the quasar nucleus using 
the C\,{\sc iv} emission line; obviously, the Ly\,$\alpha$ line
cannot be used since it appears to be absorbed, and the continuum cannot
be used because of the effects of relativistic beaming, which cannot
reliably be quantified.
The observed flux of the C\,{\sc iv} line is
$1.0\cdot10^{-14}\,$erg\,s$^{-1}$\,cm$^{-2}$.
A comparison sample can be constructed from the
optically selected submm emitting quasars studied by 
Priddey et al.\ (\cite{Priddey}), using the spectra from 
Hagen et al.\ (\cite{Hagen}). This comparison sample covers redshifts
from 2.60 to 2.79 and can therefore be compared directly to 
SMM\,J04135+10277. The comparison sample has C\,{\sc iv} fluxes from 1.7
to $4.1\cdot10^{-14}\,$erg\,s$^{-1}$\,cm$^{-2}$, roughly a factor of 3
higher than SMM\,J04135+10277.
The rest-frame {\it equivalent width\/} of C\,{\sc iv} on the
other hand shows the opposite trend: while SMM\,J04135+10277 has a
C\,{\sc iv} restframe equivalent width of $\sim 170\,$\AA, values in the
comparison sample are approximately a factor of 10 lower, ranging from
13 to $25\,$\AA. 
In other words, the quasar continuum is fainter by about a factor of 30
than would be expected for its C\,{\sc iv} flux.
It is highly unlikely that extinction could account for this, since the quasar
continuum and the broad line region should be viewed through
approximately the same obscuring column. 
Furthermore, as Fig.~\ref{fig:spectrum} shows, the quasar continuum is
characterized by a blue power law. The slope of this continuum does not
indicate the presence of abnormally large absorption. 
Therefore a more likely
explanation of the optical faintness of this quasar is a large viewing
angle away from the direction of relativistic beaming. The beamed
flux density is proportional to $\delta^p$ with $p\sim4$, where the
Doppler factor $\delta=[\gamma(1-\beta\cos\theta)]^{-1}$, where $\beta$ is
the bulk velocity in units of the speed of light, and
$\gamma=(1-\beta)^{-1/2}$ is the corresponding Lorentz factor, and
$\theta$ is the angle away from the beam
(Urry \& Padovani \cite{UrryPadovani}). 
Therefore a decrease in $\delta$ of a factor 2.3
would be sufficient to produce a factor 30 decrease in the beamed
continuum with respect to the lines. The required angle away from the
beam cannot be calculated since $\beta$ is not known. However, as shown
by Urry \& Padovani (\cite{UrryPadovani}), variations in $\delta$ of this
magnitude are entirely reasonable for angles $\theta<20\degr$, provided
$\gamma>2$. This estimate confirms the viability of our suggestion that
the optical faintness
of the quasar is due to a large viewing angle away from the direction of
relativistic beaming, and not to abnormally large extinction.
If in fact the optical spectrum is still dominated by the doppler boosted
jet then our detection of this one object suggests that a much larger 
number of yet unidentified sources are similar AGNs viewed from a larger
angle to the jet axis. 
We finally note that 
it would be interesting to make the same comparison with low-$z$ far-IR
detected quasars, addressing also the properties of the dust emssion
spectrum. This comparison would require spectrophotometry of quasars in
the vacuum ultraviolet.

Going further, we can investigate whether the observed
submm emission from SMM\,J04135+10277 is likely powered by the AGN or
whether the presence of an additional power source is indicated. In the
comparison sample, the observed $850\,\mu$m fluxes range from 6.8
to 10.0\,mJy, increasing monotonically with C\,{\sc iv} flux. The three
times fainter C\,{\sc iv} flux of SMM\,04135+10277 thus would suggest an
AGN-powered $850\,\mu$m flux of approximately 3\,mJy. The observed flux
is almost a factor of 10 higher. This result suggests that the bulk of
the submm emission from SMM\,04135+10277 is not powered by the AGN but
by an additional source of energy, most likely vigorous star formation
in the host galaxy. If this interpretation is correct, 
high resolution imaging of CO lines and dust emission with ALMA should
reveal an extended source.

\subsection{Gravitational magnification}

The low redshift ($z=0.088$) of A478 is far from the optimal lensing
redshift ($z \sim 0.2$), and no arcs are detected in the vicinity of the
quasar.  This suggests that the gravitational magnification is small
and that the quasar is not subject to differential lensing, 
which otherwise would influence the shape of the spectral energy
distribution (SED).
We calculate the magnification of the quasar using LENSTOOL (Kneib et al.\
\cite{lenstool}).  The cluster is modelled using two
components: the overall cluster potential with a core radius $r_{\rm c} =
250\ {\rm kpc}$ and a velocity dispersion $\sigma = 905\ {\rm km\,s}^{-1}$ and
the potential of the cD galaxy with $r_{\rm c} = 2\ {\rm kpc}$ and
$\sigma = 350\ {\rm km\,s}^{-1}$ (Allen et al. \cite{A478numbers1}; Zabludoff
et al. \cite{A478numbers2}).  We find that the quasar is magnified by a
factor of 1.3.  Hence, all fluxes should be corrected for this value.
In all calculations in the following sections of this 
paper the fluxes have been
corrected for the gravitational lensing.

\subsection{Spectral energy distribution}

From archival data from the NRAO Very Large Array 
a radio source near the position of the SMM\,J04135+10277 was 
found.  The fluxes measured are $S(4.86{\rm GHz})= 220\pm35\,\mu{\rm Jy}$
and $S(1.4{\rm GHz}) < 750\,\mu{\rm Jy}$ ($3\sigma$) (M.~Yun, private comm.). 
Based on this low radio flux, it is concluded that the quasar is
radio-quiet (according to the radio-power criterion given by
Stocke et al.\ \cite{stocke} to divide quasars into radio-loud and
radio-quiet types). 
Hence, the non-thermal contribution to the submm flux is expected to be 
small and is here neglected. 

Combined, we have then six points on the SED 
and two upper limits, ranging from the radio to the optical regime
(see Table~\ref{tab:flux_pos}).  
The SCUBA points and the two ISO points at 
$14.3\,\mu{\rm m}$ and $170\,\mu{\rm m}$ are in the rest frame all at 
wavelengths typical for thermal emission by dust.
The ISO $6.7\,\mu{\rm m}$
point in rest frame is $1.7\,\mu{\rm m}$, which together with the optical 
point is expected to originate from stellar light, possibly contaminated
with non-thermal emission from the AGN\null. The radio emission 
is attributed to non-thermal synchrotron emission.

In the following we will focus on the thermal dust emission. 
Of the thermal emission, the two SCUBA points are most likely due to 
the cool dust typically described by a modified blackbody, 
whereas the $14.3\,\mu{\rm m}$ point arises from a hot component. 
The shape of the SED between the cold and hot component seems to be 
different for different quasars (see e.g.\ Haas et al.\ \cite{haas_pg}). 
We have no measurements between $450\ \mu{\rm m}$ and $14.3\ \mu{\rm m}$, 
which makes an analysis of the IR SED difficult.  We do, however, attempt 
to make a tentative analysis in which we compare with known objects 
and also estimate parameters like temperature and luminosity.

\begin{figure}
\resizebox{\hsize}{!}{\includegraphics{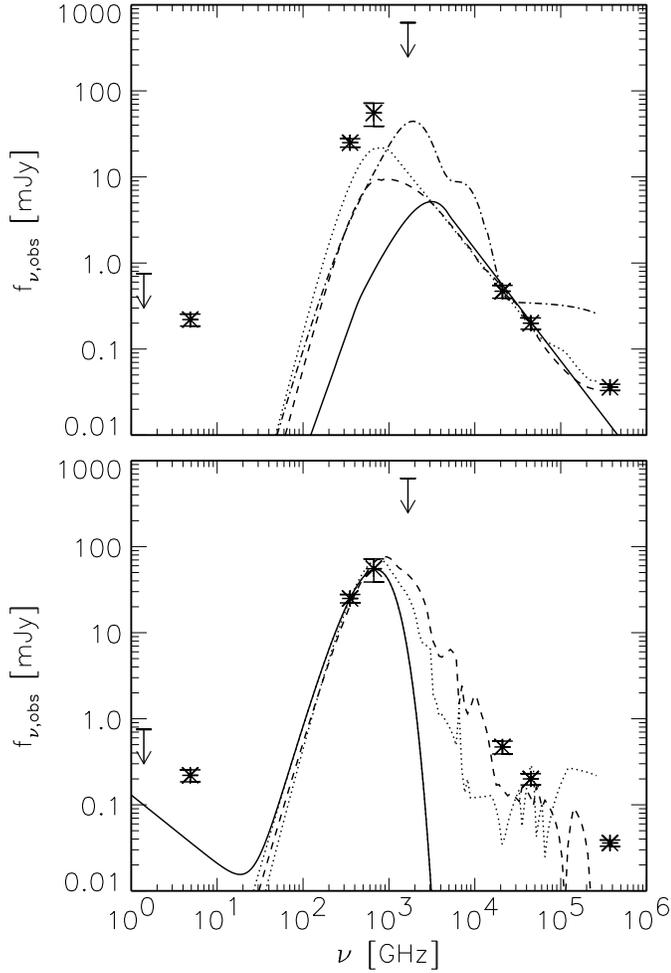}}
\caption{
The SED of the quasar SMM\,J04135+10277 compared to other SEDs.  
The SED points are given 
by the asterisks, where the error bars are $1\sigma$, and the upper limits 
are indicated by arrows. The flux densities 
as displayed in the figure have not been 
corrected for the gravitational lensing. 
As we are primarily interested in the thermal dust 
emission, the radio points have not been included in the SEDs.
{\it Upper panel:}  Comparison to the SEDs of three PG quasars (Haas et
al.\ \cite{haas_pg}):  \object{PG\,0050+124} ($z=0.061$; {\it dashed}), 
\object{PG\,1206+459} ($z=1.158$; {\it dotted}), 
and \object{PG\,1613+658} ($z=0.129$; {\it dash-dot}), and the quasar
\object{APM\,08279+5255} (z=3.87; {\it solid}; Lewis et al.\ \cite{lewis}, 
Irwin et al.\ \cite{irwin}). 
The four SEDs have been scaled to the observed $14.3\,\mu$m flux 
of SMM\,J04135+10277, as described in the text.  
{\it lower panel:} Comparison to the SEDs of the starburst galaxy NGC\,253 
({\it dotted}) and the ULIG NGC\,6240 ({\it dashed}).  Both 
SEDs have been scaled to the observed $850\,\mu{\rm m}$ flux of
SMM\,J04135+10277.   The solid line is a regular expected far-IR--radio 
correlation-based SED line. 
}
\label{fig:SED}
\end{figure}

We first compare the quasar with other known quasars.  
Comparing to high-$z$ quasars is not trivial, since the high-$z$ 
quasars
which have well-sampled IR SEDs, are often strongly lensed and their 
observed SEDs may have suffered differential lensing. We therefore
first focus on low-$z$ quasars.
Haas et al.\ (\cite{haas_pg}) have made a detailed study of the 
IR SED of Palomar-Green (PG) 
quasars. The majority of these quasars are at fairly low redshift.
We compare to
three low-$z$ PG quasars with well-sampled SEDs, \object{PG\,0050+124},
\object{PG\,1206+439} and \object{PG1613+638} (all 
shown in Fig.~\ref{fig:SED}).  All three SEDs are redshifted to $z=2.837$. 
If the SEDs are scaled to the quasar $850\,\mu{\rm m}$ point, the comparison 
gives the impression
of a deficit in the mid/near-IR emission of
SMM\,J04135+10277. Alternatively, inspired 
by the findings of Archibald et al.\ (2001) and Page et al.\ (2001), that the 
star formation rate observed in AGNs is higher at higher redshift,
leading to enhanced long-wavelength emission at higher redshift,  
we may choose instead to scale the low-$z$ SEDs to the observed 
$14.3\,\mu{\rm m}$ 
point, i.e., the hot dust emission associated with the AGN\null. 
This, not unexpectedly, then suggests an excess in the far-IR-submm 
emission from SMM\,J04135+10277. This result corroborates our earlier
conclusion that a significant portion of the observed $850\,\mu$m
emission of SMM\,J04135+10277 results from extended star formation, and
is not powered directly by the AGN\null.
The SED of the strongly lensed $z=3.87$ quasar 
APM\,08279+5255 (Lewis et al.\ \cite{lewis} and references 
therein) is also shown in Fig.~\ref{fig:SED}.  It 
has also been appropriately shifted and scaled to the $14.3\,\mu{\rm m}$
point. In this case the submm/FIR deficit relative to the 
\object{SMM\,J04135+10277} is even
more pronounced, corroborating the discussion above.

Given this result, it is also of interest to compare the SED of
SMM\,J04135+10277 to the SEDs of well-studied starburst galaxies.
We use the SEDs of the starburst galaxy \object{NGC\,253} and
the ultraluminous infrared galaxy (ULIRG) \object{NGC\,6240} 
(extracted from the NASA 
Extragalactic Database), redshifted to $z=2.837$, for comparison.
Scaled to the $850\,\mu{\rm m}$ point, the far-IR/submm
range matches quite well, whereas the mid/near-IR emission
is much brighter for SMM\,J04135+10277.  This result 
is expected, as starbursts are known not to have the hot dust
component that is characteristic of AGNs, especially quasars (Sanders et al.\ 
\cite{Sanders}; Barvainis \cite{Barvainis}). 
Considering that the SED of the quasar is not well-sampled, and that 
we are looking at only one quasar, no definite conclusions can be 
drawn about 
the precise shape of the IR SED and the power source of the dust emission
of SMM\,04135+10277 in particular, or 
of the submm-selected quasar population in general.  Tentatively, 
though, these observations support the suggestion that quasars at higher
redshift have a high submm and far-IR flux, suggesting a higher
star formation rate. 
To study this in detail, however, observations filling the 
big gaps in the IR SED are needed. Such data can possibly be
obtained with SIRTF.

\begin{figure}
\resizebox{\hsize}{!}{\includegraphics{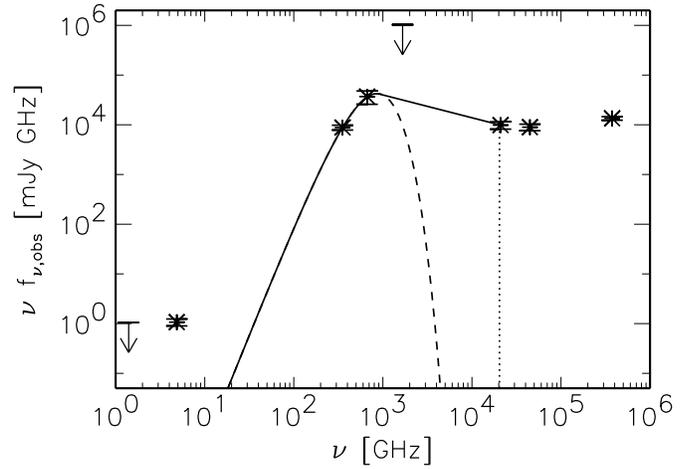}}
\caption{The quasar SED is overlayed with the SED which we have used for 
calculating the luminosity.  The flux as displayed in the figure has not 
been corrected for the gravitational magnification.}
\label{fig:QSOSED}
\end{figure}

Finally, we calculate the dust temperature, dust mass, and 
total luminosity in dust emission of
SMM\,J04135+10277. For the dust emission, we use blackbody emission,
modified by the
frequency-dependent mass absorption coefficient
\begin{equation}
k_{\rm d}(\nu_{\rm rest}) = 1.5\,{\rm cm}^2\,{\rm g}^{-1} \left(
  \frac{\nu_{\rm rest}}{375\ {\rm GHz}} \right)^\beta, 
\end{equation}
using the average value from the literature at $800\,\mu$m (see 
Hughes et al.\ 1997 for a discussion of the assumptions and uncertainties
in this parameter)
and assume $\beta = 1.5$. 
Since the IR SED cannot be fit by a single modified blackbody, we
first fit the cold dust component sampled by the SCUBA points. 
Using only 
the SCUBA points, we find a temperature $T_{\rm cold} = 29\pm 2\ {\rm K}$ and 
a dust mass of $M_{\rm d} = (1.8\pm0.3)\times
10^{9}\ {\rm M_{\odot}}$ for the cold dust component.  
The total luminosity of a modified blackbody spectrum can be 
calculated using the following analytical expression:
\begin{equation}
L_{\rm tot} = 4\pi D_{\rm L}^2 \frac{S(\nu_{\rm obs})}{\nu_0^{\beta}B(\nu_0,T)} \frac{2h}{c^2}
\left ( \frac{kT}{h} \right ) ^{4+\beta} \Gamma (4+\beta) \zeta(4+\beta),
\end{equation}
where $h$ and $k$ are Planck's respectively Boltzmann's constants,
$D_{\rm L}$ is 
the luminosity distance, 
and the two last factors are the Gamma function 
and Riemann's zeta function.  $S(\nu_{\rm obs})$ is the flux density at the 
observing frequency, and $\nu_0$ is the corresponding rest frequency.
The total luminosity of the cold component 
is found to be $(2.4\pm0.5) \cdot 10^{13}\,{\rm L_\odot}$. 
To estimate the total 
IR luminosity, we assume a powerlaw between the peak of the modified 
blackbody curve and the observed $14.3\,\mu{\rm m}$ point.  The powerlaw is 
integrated from $200\,\mu{\rm m}$ (observed frame), 
where the modified blackbody and the 
powerlaw balance eachother, to $14.3\,\mu{\rm m}$ with the result of
$(5.4\pm 1)\cdot 10^{12}\,{\rm L_\odot}$.
In total the IR luminosity (corrected for gravitational amplification) 
is then $L_{\rm IR} = (2.9\pm 0.5)\cdot 10^{13}\,{\rm L_\odot}$, dominated by
the cold dust component.
Using a different method based on the analysis in Blain et al.\ 
(\cite{blain03}), where the whole IR SED is fit with a single temperature
modified blackbody with a powerlaw on the Wien side 
ranging all the way into the mid-IR, a temperature of 38\,K is found and a 
total IR luminosity of $1.8\times 10^{13}\,{\rm L_{\odot}}$.  This gives 
a higher temperature, though a slightly lower luminosity, compared to the 
fit above where a cold component was fitted to the two SCUBA points. 

The temperature as we find is lower than that found in
other high-$z$ quasars
such as APM\,08279+5255, which has a temperature of $120-220\,$K determined
for a pure blackbody (Lewis et al.\ \cite{lewis}), or BR\,1202$-$0725, 
which has 
a dust temperature of $50-68\,$K (Leech et al.\ \cite{leech}).  
Both quasars have
luminosities in order of $10^{14-15}\,{\rm L_{\odot}}$, thus brighter than 
SMM\,J04135+10277, so that higher dust temperatures might be
expected. On the other hand, these two quasars are strongly lensed and
it is possible that differential lensing distorts the integrated SED and
overemphasizes warm dust components.


We finally attempt to compare the radio-submm flux density ratio with the 
relevant simulations performed by Blain (\cite{blain}), which are based
on the IR-radio correlation observed at low redshift.  As we do not 
have a 1.4\,GHz flux density measurement, we estimate it by assuming 
that the radio SED is a power law, $f\propto \nu^{-\alpha}$, with slope
$\alpha = -0.8$ and scale it to the observed 
flux density at 4.86\,GHz.  We find
$f_{1.4\,{\rm GHz}} = 595\,\mu{\rm Jy}$ (not corrected for the gravitational 
lensing).  Still assuming $\beta = 1.5$, 
we use Fig.~4 in Blain (\cite{blain}) by interpolating between his two 
models with $T=20\,{\rm K}$ respectively $T=40\,{\rm K}$.  For $z=2.837$ 
this gives a flux density ratio of between $1.4\,$GHz and $850\,\mu$m of 
$\sim3.5\cdot10^{-3}$. Therefore, the observed $850\,\mu$m flux would
imply 
$f_{1.4\,{\rm GHz}} \sim 88\,\mu{\rm Jy}$ if the quasar strictly
followed the local IR-radio correlation. This number is however a factor $6-7$ 
lower than what we had just estimated based above.  This indicates that 
SMM\,J04135+10277 has more radio emission (for its IR emission) 
than e.g., the ULIRG 
\object{Arp220}, 
which was used for the template SED in Blain (1999).  This result is not 
surprising, as quasar radio emission is powered by both the synschrotron
emission from stellar remnants and the synchrotron emission from the 
central black hole. 


\section{Conclusions}
\label{sect:concl}

We have discovered a type-1 quasar behind the cluster of galaxies A478.  
The quasar, \object{SMM\,J04135+10277}, was discovered by it submm emission.  
The quasar has a redshift of $z=2.837$ and is radio quiet.
The quasar is optically faint, but 
has a large submm flux. Using number counts of quasars and of
submm sources, we argue that SMM\,04135+10277 is an unusual
object. It is in any case a remarkable object 
since there is little overlap between
deep submm and X-ray samples, suggesting that the incidence of 
powerful AGNs among submm galaxies is low.
The slope of the rest-frame UV continuum is 
similar to that of optically selected quasars 
(Francis et al.\ \cite{Fran92}) 
and does not exhibit any signs of extraordinary dust
extinction.  This leads us to suggest that the line of sight to the
quasar nucleus is not abnormally
obscured. Analysis of optical continuum, spectral lines and submm
emission leads us to conclude that the optical faintness of the quasar
results from a large viewing angle from the direction of relativistic
beaming, and that a significant amount of the submm flux is not powered
by the active nucleus. 
More likely the cold dust is heated by a high rate of star
formation in the environment surrounding the quasar. Comparison of the
sparsely sampled
IR SED to that of other objects tentatively supports this conclusion.
The total IR luminosity is found to be $(2.9\pm0.5)\cdot
10^{13}\,{\rm L_{\odot}}$ and is dominated by the emission from cool dust.


\begin{acknowledgements}
  
  We thank Remo Tilanus for taking most of the SCUBA data presented in
  this paper, Min Yun for providing us with the VLA archive data, and 
  Jean-Paul Kneib for making his LENSTOOL program available for us.
  We also thank the referee, Andrew Blain, for useful comments. 
  KKK is supported by the Netherlands Organization for Scientific
  Research (NWO).  The JCMT is operated by the Joint Astronomy Centre on
  behalf of the United Kingdom Particle Physics and Astronomy Research
  Council (PPARC), the Netherlands Organization for Scientific Research,
  and the National Research Council of Canada.  The National Radio
  Astronomy Observatory is a facility of the National Science
  Foundation, operated under cooperative agreement by Associated
  Universities, Inc.  This research has made use of the NASA/IPAC
  Extragalactic Database (NED) which is operated by the Jet Propulsion
  Laboratory, California Institute of Technology, under contract with
  the National Aeronautics and Space Administration.  The ISOCAM data
  presented in this paper was analysed using ``CIA'', a joint
  development by the ESA Astrophysics Division and the ISOCAM
  Consortium. The ISOCAM Consortium is led by the ISOCAM PI,
  C.~Cesarsky.  The ISOPHOT data presented in this paper was reduced
  using PIA, which is a joint development by the ESA Astrophysics
  Division and the ISOPHOT consortium, with the collaboration of the
  Infrared Analysis and Processing Center (IPAC) and the Instituto de
  Astrofísica de Canarias (IAC).

\end{acknowledgements}

\end{document}